\documentclass[hidelinks,twocolumn,showpacs,aps,prd,superscriptaddress,floatfix]{revtex4-1}
\usepackage{graphicx}
\usepackage{verbatim}
\usepackage{dcolumn}
\usepackage{amsmath,amsfonts,amssymb}
\usepackage{epsfig}
\usepackage{subfigure}
\usepackage{url}

\usepackage{breakurl}
\usepackage[breaklinks]{hyperref}

\RequirePackage{xspace}

\newcommand{\GeV}{\ensuremath{\mathrm{\,Ge\kern -0.1em V}}\xspace}
\newcommand{\MeV}{\ensuremath{\mathrm{\,Me\kern -0.1em V}}\xspace}

\def\ee         {\ensuremath{e^+e^-}\xspace}

\def\mumu       {\ensuremath{\mu^+\mu^-}\xspace}

\def\tautau     {\ensuremath{\tau^+\tau^-}\xspace}

\def\ellell     {\ensuremath{\ell^+ \ell^-}\xspace}

\def\x          {\ensuremath{\chi}\xspace}
\def\xb         {\ensuremath{\bar{\chi}}\xspace}
\def\xx         {\ensuremath{\chi\bar{\chi}}\xspace} 

\def\ra    {\rightarrow}
\def\p     {\prime}
\def\pL    {\ensuremath{\phi}\xspace}
\def\mpL   {\ensuremath{m_{\phi}}\xspace}
\def\Ap    {\ensuremath{A^{\prime}}\xspace}
\def\mAp   {\ensuremath{m_{A^{\prime}}}\xspace}

\def\figurebox#1#2#3{%
    \def\arg{#3}%
    \ifx\arg\empty
    {\hfill\vbox{\hsize#2\hrule\hbox to #2{\vrule\hfill\vbox to #1{\hsize#2\vfill}\vrule}\hrule}\hfill}%
    \else
    {\hfill\epsfbox{#3}\hfill}%
    \fi}

\begin{document}

\pagestyle{plain}

\title{{\large \bf Prospects for Dark Matter Search at the Super c-tau Factory}}

\author{E.\,E.\, Boos} \email{boos@theory.sinp.msu.ru}

\author{V.\,E.\, Bunichev} \email{bunichev@theory.sinp.msu.ru}

\author{S.\,S.\, Trykov} \email{trykow.sergei@gmail.com}

\affiliation{Skobeltsyn Institute of Nuclear Physics, Lomonosov Moscow State University}

\begin{abstract}
We present perspectives for searching for light dark matter production mediated by a leptophilic scalar~$\pL$ and a dark photon~$\Ap$ in in experiments at the~Super c-tau Factory. 
Based on the analysis of the associative production of mediators and $\tau$-leptons at the energies of the future collider, the possibility of searching in the non-excluded region of the parameter space was found.
The obtained sensitivity curves at the $90\%$ C.L. for the mediators mass range below $4\,\GeV$ demonstrate the~power of the~SCTF for light dark matter search.
\end{abstract}

\maketitle

\setcounter{footnote}{0}

\section{Introduction}

Many gravitational, astrophysical and cosmological unexplained phenomena indicate the existence of a special kind of matter, 
which we call dark matter (DM)~\cite{Zwicky:1933gu,Zwicky:1937zza,Rubin:1970zza}. 
Such kind of phenomena can be explained by assuming that DM is a kind of particles, which should have some certain properties. 
We follow a simplified models approach, in which it is assumed that there are DM particles of any spin and there are particles, 
usually scalar or vector, mediating interactions between Standard Model (SM) particles and DM~\cite{Agrawal:2021dbo}.

Many models beyond the SM predict the existence of additional scalars $\pL$ 
that can mediate interactions between SM particles and DM~\cite{Marshall:2010qi,Branco:2011iw,Batell:2016ove}. 
Possible couplings of an interaction between additional scalars with SM fermions are constrained by the SM gauge invariance. 
Now, the minimal extension of the scalar sector by mixing an additional scalar with the SM Higgs boson is strongly constrained by experiments of searching for rare flavor-changing neutral current decays of mesons, such as $B^{+}\ra K^{+}\pL$ and $K^{+}\rightarrow\pi^{+}\pL$~\cite{Beacham:2019nyx,Lanfranchi:2020crw}, and by searching for heavy DM~\cite{Felcini:2018osp,Beacham:2019nyx,Lanfranchi:2020crw,Agrawal:2021dbo}. 
Thus, the~$\MeV-\GeV$ range search area comes to the fore.

In a sub-GeV scale, some BSM theories make it possible to assume that in the SM sector after spontaneous electroweak symmetry breaking, an extra scalar couples exclusively to SM leptons~\cite{Marshall:2010qi,Branco:2011iw,Batell:2016ove}. 
Possible couplings with quarks turn out to be strongly suppressed.
Thus, a scalar portal for interaction between heavy-flavored SM leptons and DM is occurred. 
We refer the additional scalar particle as a dark leptophilic scalar $\pL$. 
A Lagrangian of that interaction is given by
\begin{equation}\label{L-pL}
\mathcal{L}_{\rm int}^{\pL}=-\xi\sum\limits_{\ell=e,\mu,\tau}\frac{m_{\ell}}{v}\bar{\ell}\pL\ell-g_{\rm D}\xb\pL\x,
\end{equation}
where $\xi$ is the flavor-independent coupling constant, $v=246\,\GeV$ is the SM Higgs vacuum expectation value, and the second term is a Lagrangian of the interaction between the dark scalar $\pL$ and fermionic DM states $\x$ with a coupling constant $g_{\rm D}$.

In the~$\MeV-\GeV$ mass range, if kinematically allowed, dominant decay modes of the dark leptophilic scalar are to lighter DM states and to leptons, 
with partial widths given by~\cite{Abercrombie:2015wmb}
\begin{align}
\Gamma_{\pL}^{\xx}		&=g_{\rm D}^2\frac{m_{\pL}}{8\pi}\beta_{\chi}^3, \label{w-pL-xx} \\
\Gamma_{\pL}^{\ellell}	&=\xi^2\frac{m_{\ell}^2}{v^2}\frac{m_{\pL}}{8\pi}\beta_{\ell}^3, \label{w-pL-ll}
\end{align}
here $\beta_f=\sqrt{1-4 m_f^2/m_{\pL}^2}$, ~$m_{\pL}$ - mass of the mediator, ~$m_f$ - mass of the mediator decay product.

Another intriguing possibility for providing a portal between the SM and DM sectors is an addition of an extra massive vector boson associated with the spontaneously broken $U_{\rm D}(1)$ gauge group, 
whose interaction between the SM charged fermionic current is similar to the ordinary photon of electromagnetism~\cite{Holdom:1985ag,Alexander:2016aln,Agrawal:2021dbo}. 
We refer such a vector mediator as a dark photon $\Ap$. 
The $\Ap$ can have a mass in the sub-GeV mass range, and couple to the SM via kinetic mixing with the ordinary photon.
A coupling constant of the interaction between $\Ap$ and the SM states is suppressed by a kinetic mixing parameter $\varepsilon\ll 1$. 
A Lagrangian can be written as~\cite{Okun:1982xi,Galison:1983pa,Holdom:1985ag}
\begin{equation}\label{L-Ap+xx}
\mathcal{L}=\mathcal{L}_{\rm SM}+\frac{1}{2}\frac{\varepsilon}{\cos\theta_{\rm {W}}}B^{\mu\nu}F^{\p}_{\mu\nu}+\mathcal{L}_{\rm Dark}
-e_{\rm D}\Ap_{\mu}\,j_{\rm DM}^{\mu},
\end{equation}
where $F^{\p}_{\mu\nu}\equiv \partial_{\mu}\Ap_{\nu}-\partial_{\nu}\Ap_{\mu}$ is the dark photon gauge field $\Ap_{\mu}$ strength tensor,  
$B_{\mu\nu}\equiv \partial_{\mu} B_{\nu}-\partial_{\nu} B_{\mu}$ is the SM weak hypercharge strength tensor;
$e=\sqrt{4\pi\alpha_{\rm EM}}$ is the $U(1)$ coupling constant, and
$e_{\rm D}=\sqrt{4\pi\alpha_{\rm D}}$ is the $U_{\rm D}(1)$ coupling constant of the interaction between $\Ap$ and the DM current~$j^{\mu}_{\rm DM}=\xb\gamma^{\mu}\x$ for fermionic DM states.
$\mathcal{L}_{\rm Dark}$ is a term containing a DM and dark photon states Lagrangians.

After spontaneous symmetry breaking in the $\MeV-\GeV$ mass range, 
the main contribution of mixing by the term $B^{\mu\nu}F^{\prime}_{\mu\nu}$ in the Lagrangian \eqref{L-Ap+xx} is given by 
$(\varepsilon/2)F^{\mu\nu}F^{\prime}_{\mu\nu}$. 
Mixing with a heavy $Z$ boson is suppressed by a factor $1/m_{Z}^2$. 
After diagonalization of the kinetic terms, the result of mixing leads 
to the fact that the dark photon $\Ap$ acquires a coupling $\varepsilon e$ 
with the electromagnetic current $j^{\mu}_{\rm EM}$:
\begin{equation}\label{L-Ap}
\mathcal{L}_{\rm int}^{\Ap}=-\varepsilon e\Ap_{\mu}\,j_{\rm EM}^{\mu}-e_{\rm D}\Ap_{\mu}\,j_{\rm DM}^{\mu}.
\end{equation}

We consider light DM states $\x$ with mass $m_{\x}<m_{\Ap}/2$ and $e_{\rm D}>e$.
Thus, if kinematically allowed, dark photon are expected to decay predominantly into invisible dark sector final states.
If no such decays are allowed, the dark photon will decay into visible SM final states, with decay partial widths given by~\cite{Abercrombie:2015wmb,Fabbrichesi:2020wbt}
\begin{equation}
\Gamma_{\Ap}^{\xx}=\frac{1}{3}\alpha_{\rm D} m_{\Ap}
\left(1+2\frac{m_{\chi}^2}{m_{\Ap}^2}\right)\beta_{\x},
\end{equation}
\begin{align}
\Gamma_{\Ap}^{\ellell}		&=\frac{1}{3}\varepsilon^2\alpha m_{\Ap}
\left(1+2\frac{m_{\ell}^2}{m_{\Ap}^2}\right)\beta_{\ell}, \\
\Gamma_{\Ap}^{\rm hadrons}	&=\frac{1}{3}\varepsilon^2\alpha m_{\Ap}
\left(1+2\frac{m_{\mu}^2}{m_{\Ap}^2}\right)\beta_{\mu}\,R, \\
R							&\equiv\frac{\sigma(\ee\ra{\rm hadrons})}{\sigma(\ee\ra\mumu)}.
\end{align}
where $\beta_f$ has the same definition as in \eqref{w-pL-xx} and \eqref{w-pL-ll}. 

\begin{figure}[htb]
\begin{center}
\includegraphics[width=0.24\textwidth]{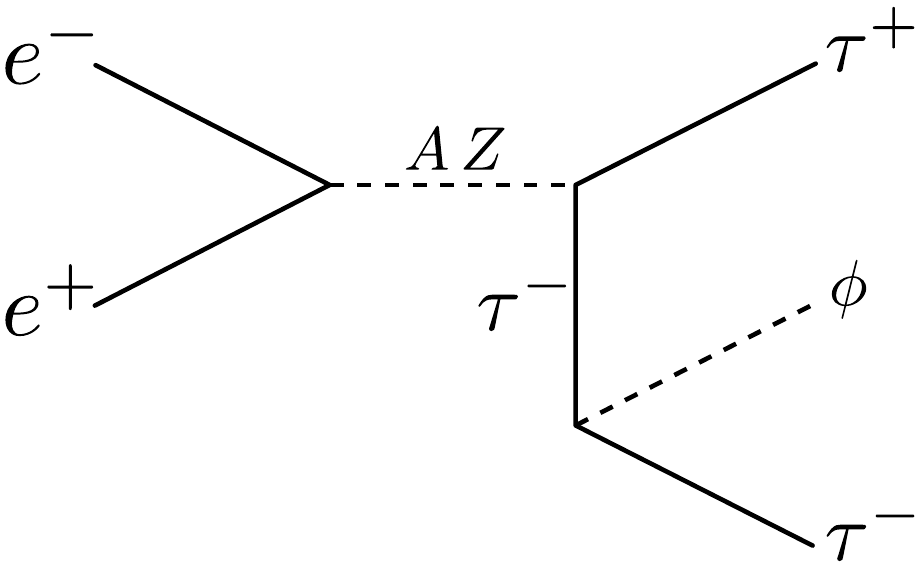}
\includegraphics[width=0.48\textwidth]{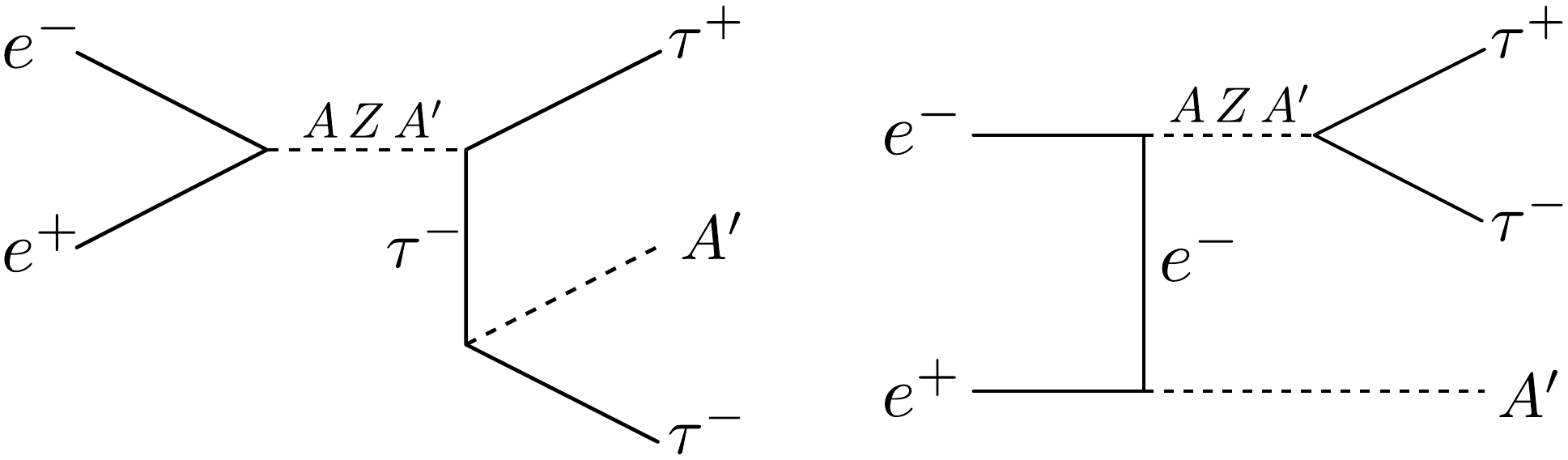}
\end{center}
\caption{The Feynman diagrams for the $\pL$ and $\Ap$ on-shell production in $\ee$ collisions. 
We assume that the~mediators subsequently decays to lighter DM.}
\label{feynman}
\end{figure}

In this work we focus on searching for invisible decays of the leptophilic scalar~$\pL$ and the dark photon~$\Ap$ in the processes 
$\ee\ra\tautau +(\pL\ra invisible)$ and $\ee\ra\tautau +(\Ap\ra invisible)$, which are shown in the Fig.~\ref{feynman}, 
at the running energies of the future Super c-tau Factory (SCTF).
The~SCTF is a promising project of a future electron-positron collider for the energy range from $3$ to $7\,\GeV$ in the centre of mass system 
that will provide a uniquely high peak luminosity of $10^{35}\,{\rm cm}^{-2}{\rm s}^{-1}$.
The~SCTF physical program is aimed at a detailed study of the processes involving the $c$-quark and $\tau$-lepton in the final state.
For $10$ years of SCTF operation, the integral luminosity of the collider will exceed $10\,{\rm ab}^{-1}$.

\section{Cross Sections}

\begin{figure*}[htb]
\begin{center}
\includegraphics[width=0.45\textwidth]{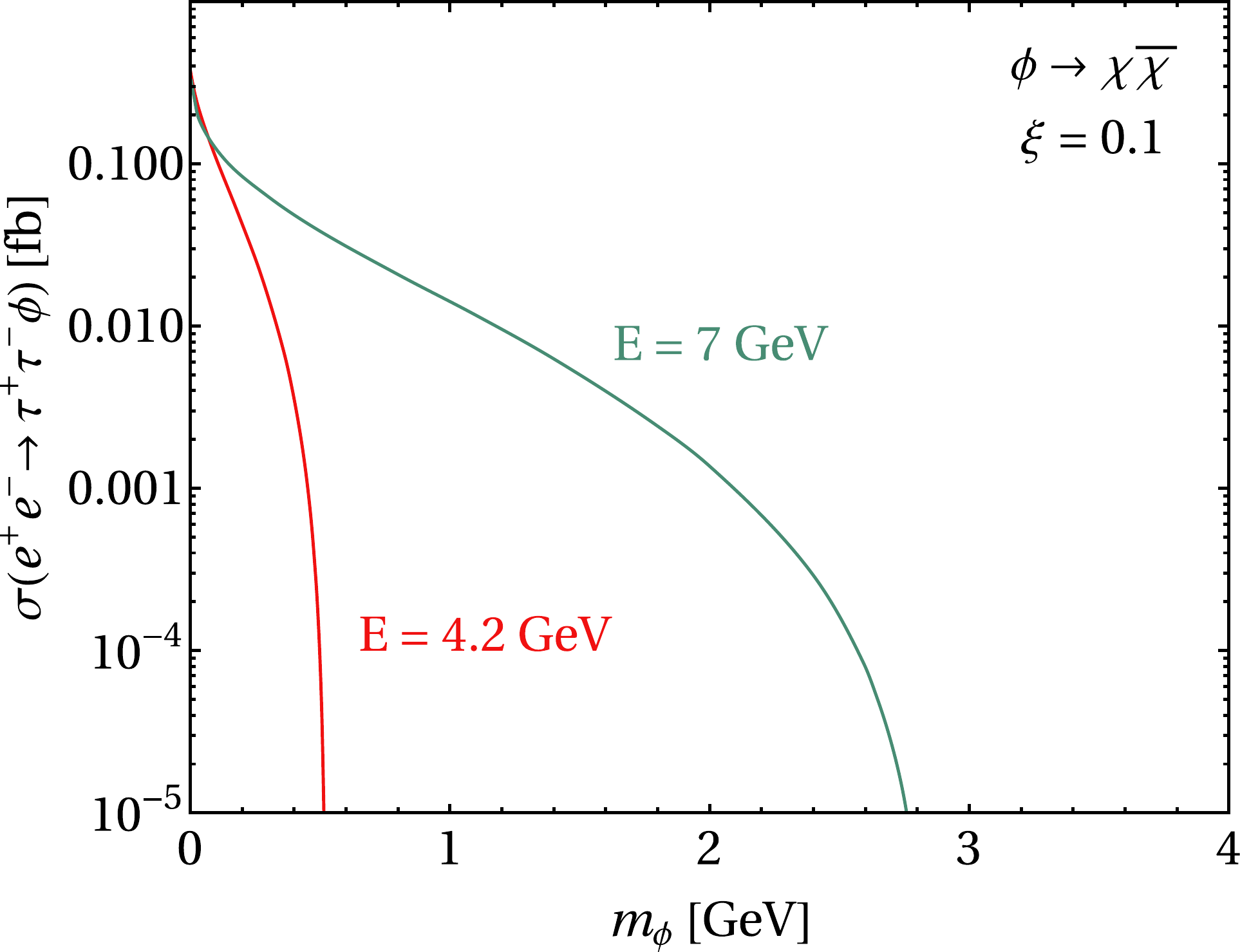}
  \hspace{1cm}
\includegraphics[width=0.45\textwidth]{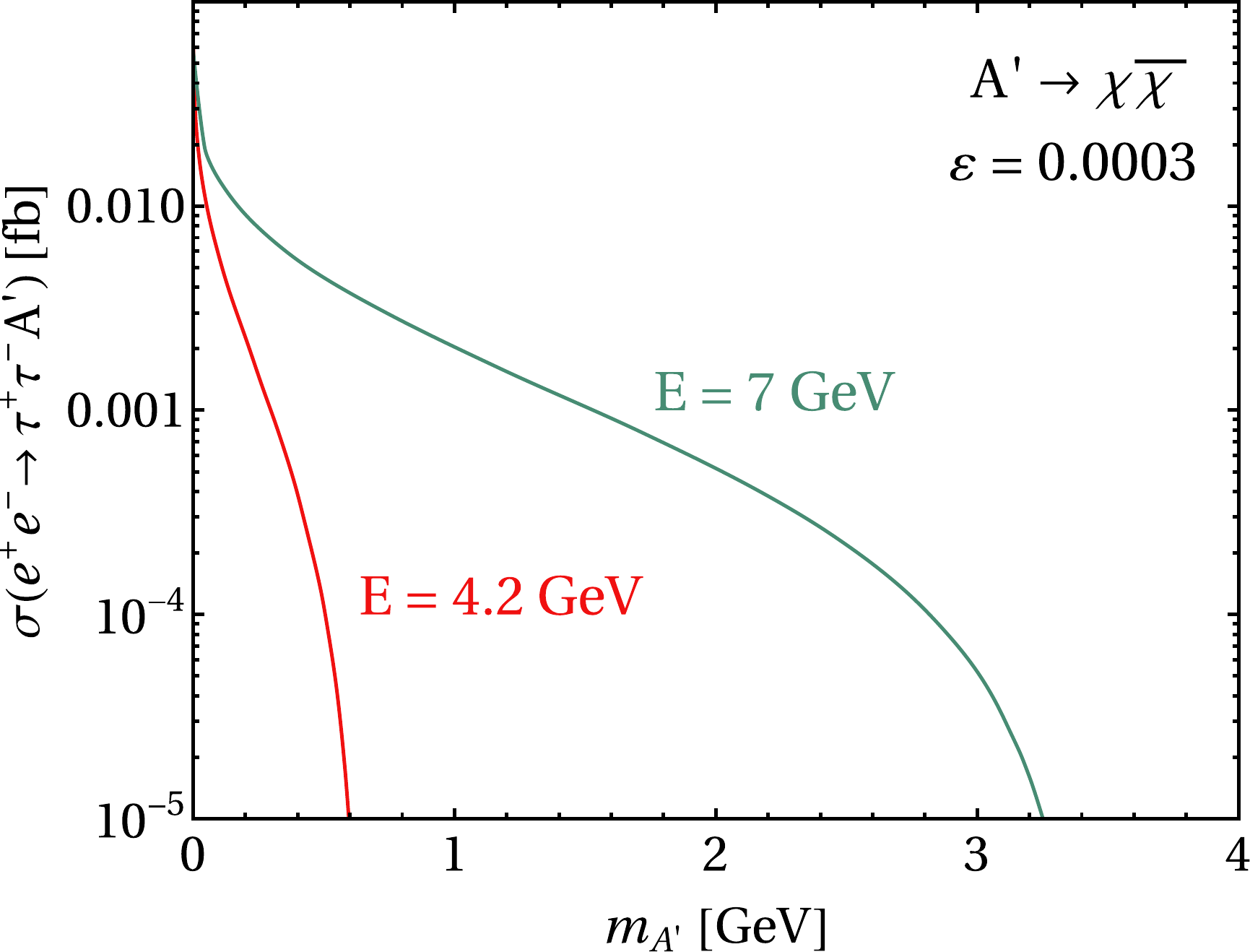}
\end{center}
\caption{The dependencies of $\ee\ra\tautau+(\pL\ra invisible)$ [left panel] and $\ee\ra\tautau+(\Ap\ra invisible)$ [right panel] cross sections on the mediators mass at the running energies of the future SCTF.}
\label{med-cross-sec}
\end{figure*}

All calculations and Monte Carlo simulation for signal and background processes were performed using the CompHEP package~\cite{CompHEP:2004qpa}. In the following we assume that the mediators invisible decay mode is predominant, $Br(\pL\ra\xx)\simeq 1$ and $Br(\Ap\ra\xx)\simeq 1$. 
If such invisible $\pL$ and $\Ap$ exists, they could be produced
by $\ee$ collisions at the SCTF and generate a flux of DM particles, which can
be detected through the missing energy and momentum.
We found that, within the framework of the considered models, the optimal collider regimes for searching for dark matter are sessions at energies of $4.2\GeV$ and $7\,\GeV$. To accurately take into account radiation effects, we used the following planned beam parameters:
the~radii of the bunches in the horizontal
and vertical dimensions are $\sigma_x=17.8\,{\rm \mu m}$ and $\sigma_y=0.178\,{\rm \mu m}$, respectively, with bunch length $\sigma_z=10\,{\rm mm}$  at the interaction point,
the number of particles per bunch $N_b=7.1\times 10^{10}$.
In Fig.~\ref{med-cross-sec}, we provide the $\pL$ and $\Ap$ production cross-section dependencies on their masses
at the~energies of $4.2$ and $7\,\GeV$, taking into account the NLO corrections from bremsstrahlung and ISR~\cite{CompHEP:2004qpa,Berends:1982dy,Berends:1987cd,Berends:1989zr,Jadach:1999vf,Akers:1995er,Banerjee:2007is}.

The main SM background to the $\ee\ra\tautau\pL$ and $\ee\ra\tautau\Ap$ signals at the future collider are the processes with a similar signature with missing energy
$\ee\ra\tautau\bar{\nu}_{\ell}\nu_{\ell}$, where $\nu_{\ell}=\nu_{e},\nu_{\mu},\nu_{\tau}$ are the SM neutrinos.
For the electron-positron collider running with $\GeV$ beam energy
such processes are suppressed by $Z$- and $W$-bosons propagators.

For each collider mode, using the statistical approach described in \cite{Bityukov:2008zz}, we estimate $90\%$ C.L. areas in the model parameter space available for research at the SCTF.

\section{Dark Leptophilic Scalar}

\begin{figure}[htb]
\begin{center}
  \includegraphics[width=0.48\textwidth]{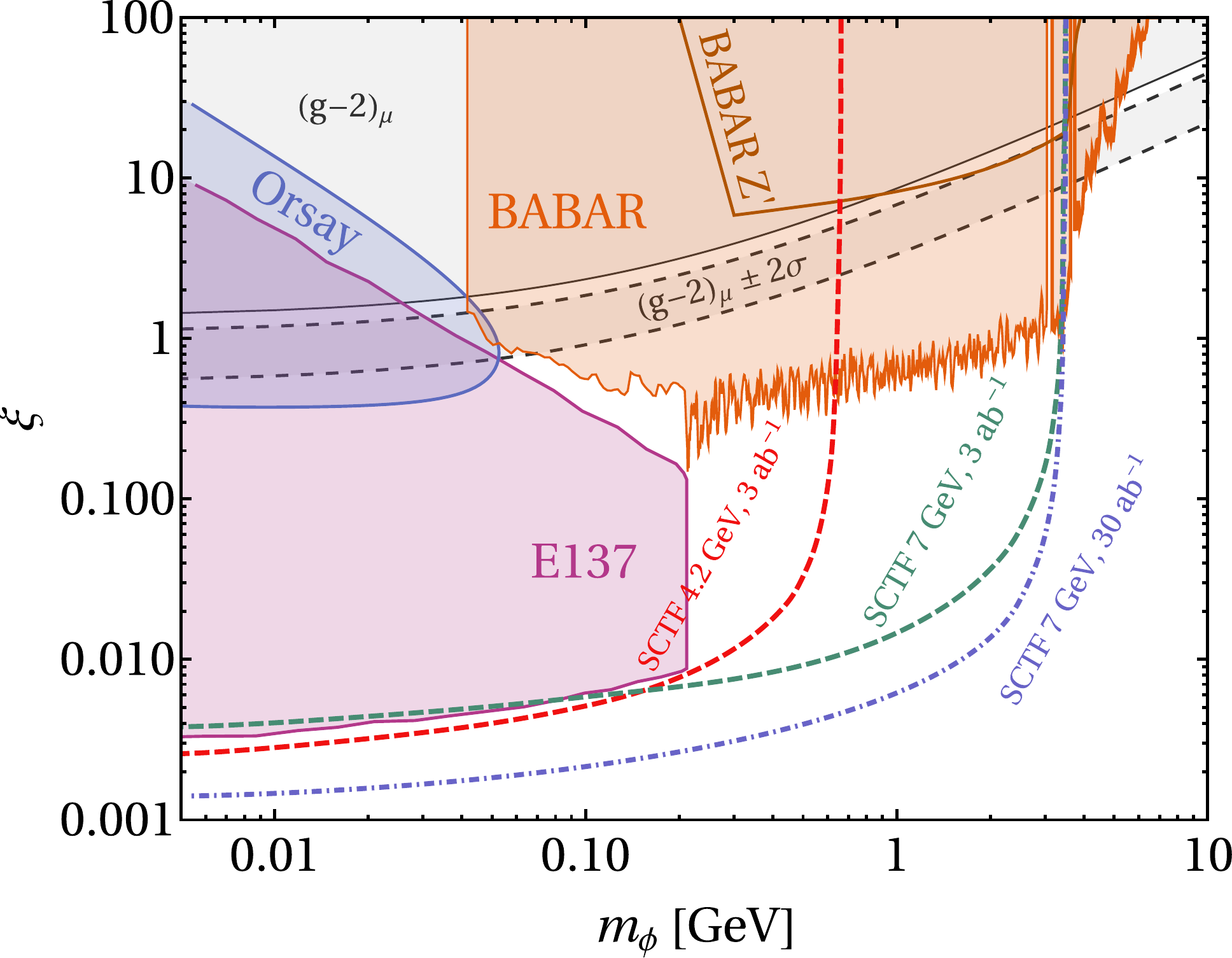}
\end{center}
\caption{The Sensitivity curves on the coupling $\xi$ as a function of the $\pL$ mass at the $90\%$ C.L. are obtained assuming 
integrated luminosity of $3\,{\rm ab}^{-1}$ at the collider energies $\sqrt{s}=4.2\,\GeV$ (dashed red), $\sqrt{s}=7\,\GeV$ (dashed green), and assuming integrated luminosity of $30\,{\rm ab}^{-1}$ at $\sqrt{s}=7\,\GeV$ (dashed blue).
Existing constraints~\cite{BaBar:2020jma,Liu:2020qgx,BaBar:2016sci,Liu:2016qwd,Davier:1989wz,Bjorken:1988as} (shaded areas) 
as well as the favored muon anomalous magnetic moment $(g-2)_{\mu}$ area~\cite{Batell:2016ove,Liu:2020qgx} (gray dashed) are also shown.}
\label{pL}
\end{figure}

First, we estimated the available parameter ranges for models with a scalar mediator. We report new results for sensitivity to the dark leptophilic scalar production with its subsequently decay to light DM in experiments at the SCTF. 
Fig.~\ref{pL} shows the sensitivity curves at the $90\%$ C.L. in the~$[\xi,\mpL]$ plane, assuming about 
$3\,{\rm ab}^{-1}$ data collected at $\sqrt{s}=4.2\,\GeV$ (dashed red), $\sqrt{s}=7\,\GeV$ (dashed green) and 
$30\,{\rm ab}^{-1}$ at $\sqrt{s}=4.2\,\GeV$ (dashed blue)
The existing experimental constraints are also shown: 
bounds in channels where $\pL$ is allowed to decay visibly from the BABAR experiment at SLAC~\cite{BaBar:2020jma},
the measurement for $Br(K_{\rm L}\ra\pi^0\pL)$ from the KOTO experiment~\cite{Liu:2020qgx}, 
search for dark bosons $Z^{\p}$ with vector couplings only to the second and third generations of leptons~\cite{BaBar:2016sci}, 
measurement from~\cite{Liu:2016qwd},
from the electron beam-dump experiment~\cite{Davier:1989wz},
search for neutral objects on the SLAC beam dump~\cite{Bjorken:1988as}.
It can be seen that the boundaries of the regions accessible to SCTF are much lower and wider than those obtained in previous experiments. Due to the high integral luminosity, SCTF allows one to "feel" mediators with a range of masses below $4\,\GeV$ and coupling $\xi$ down to $10^{-3}$.

\section{Dark Photon}

\begin{figure}[htb]
\begin{center}
  \includegraphics[width=0.48\textwidth]{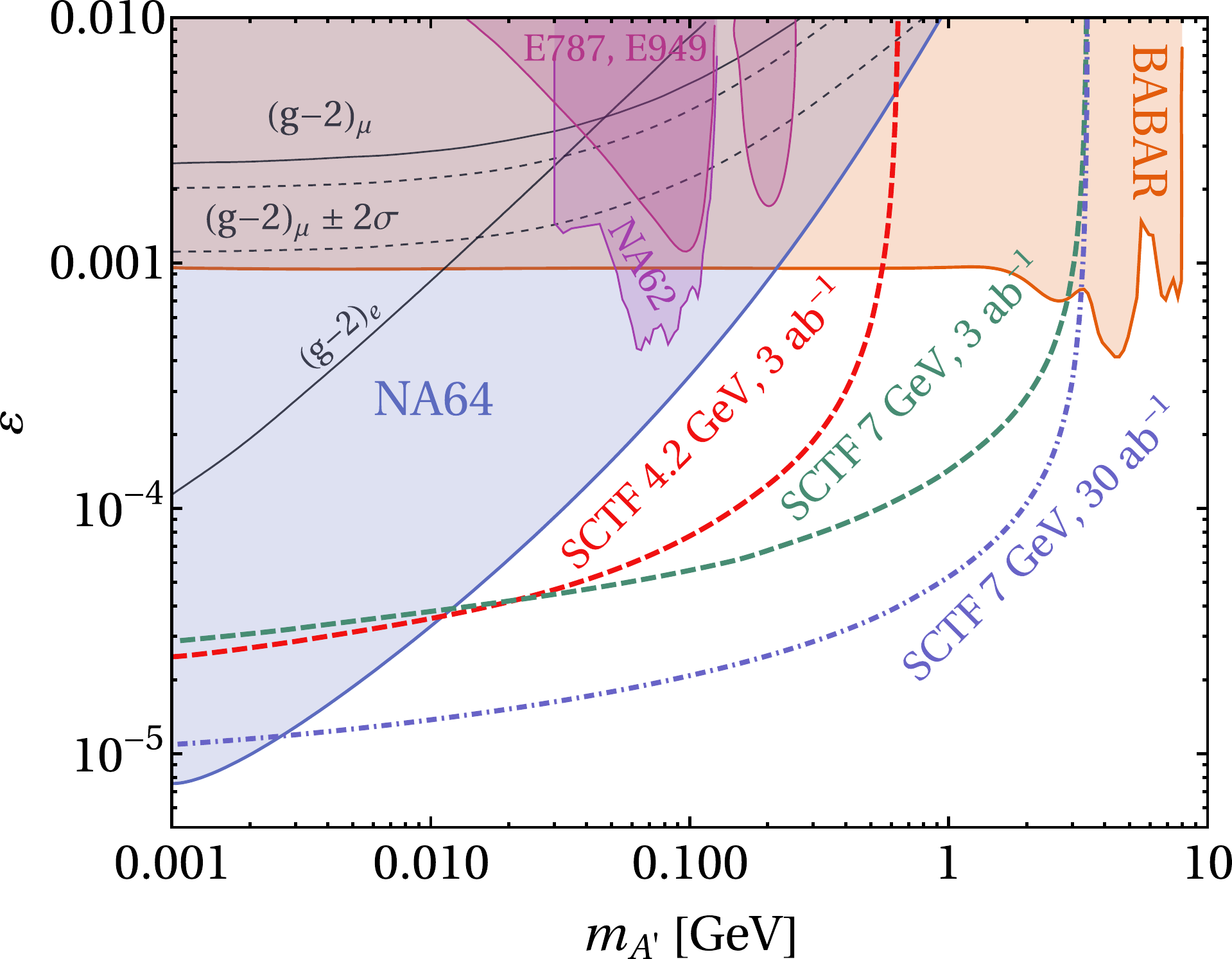}
\end{center}
\caption{The Sensitivity curves on the mixing strength $\varepsilon$ as a function of the $\Ap$ mass at the $90\%$ C.L. are obtained assuming 
integrated luminosity of $3\,{\rm ab}^{-1}$ at the collider energies $\sqrt{s}=4.2\,\GeV$ (dashed red), $\sqrt{s}=7\,\GeV$ (dashed green), and assuming integrated luminosity of $30\,{\rm ab}^{-1}$ at $\sqrt{s}=7\,\GeV$ (dashed blue).
Existing constraints~\cite{Banerjee:2019pds,NA62:2019meo,BaBar:2017tiz,Davoudiasl:2014kua,Essig:2013vha,E787:2001urh,BNL-E949:2009dza} (shaded areas) 
as well as the favored muon anomalous magnetic moment $(g-2)_{\mu}$ area~\cite{Muong-2:2006rrc} (gray dashed) are also shown.}
\label{Ap}
\end{figure}
On the next step we estimated the available parameter ranges for models with a vector mediator.
In Fig.~\ref{Ap} we present the SCTF sensitivity curves at the $90\%$ C.L. in the $[\varepsilon,\mAp]$ plane. 
The existing experimental constraints are also shown: 
bounds in channels where $\Ap$ is allowed to decay invisibly from the NA62~\cite{NA62:2019meo}, NA64~\cite{Banerjee:2019pds}, BABAR~\cite{BaBar:2017tiz},
the measurement for $Br(K^{+}\ra \pi^{+}\nu\bar{\nu})$ by the E787~\cite{E787:2001urh} and E949~\cite{BNL-E949:2009dza} experiments, 
as well as the anomalous muon magnetic moment $(g-2)_{\mu}$ favored area~\cite{Muong-2:2006rrc}.

We can see that the SCTF with about $30\,{\rm ab}^{-1}$ can provide new data in the non-excluded region 
$0.01\,\GeV\lesssim \mAp \lesssim 3.5\,\GeV$ and $\varepsilon$ down to $10^{-5}$.

\begin{figure*}[htb]
\begin{center}
  \includegraphics[width=0.48\textwidth]{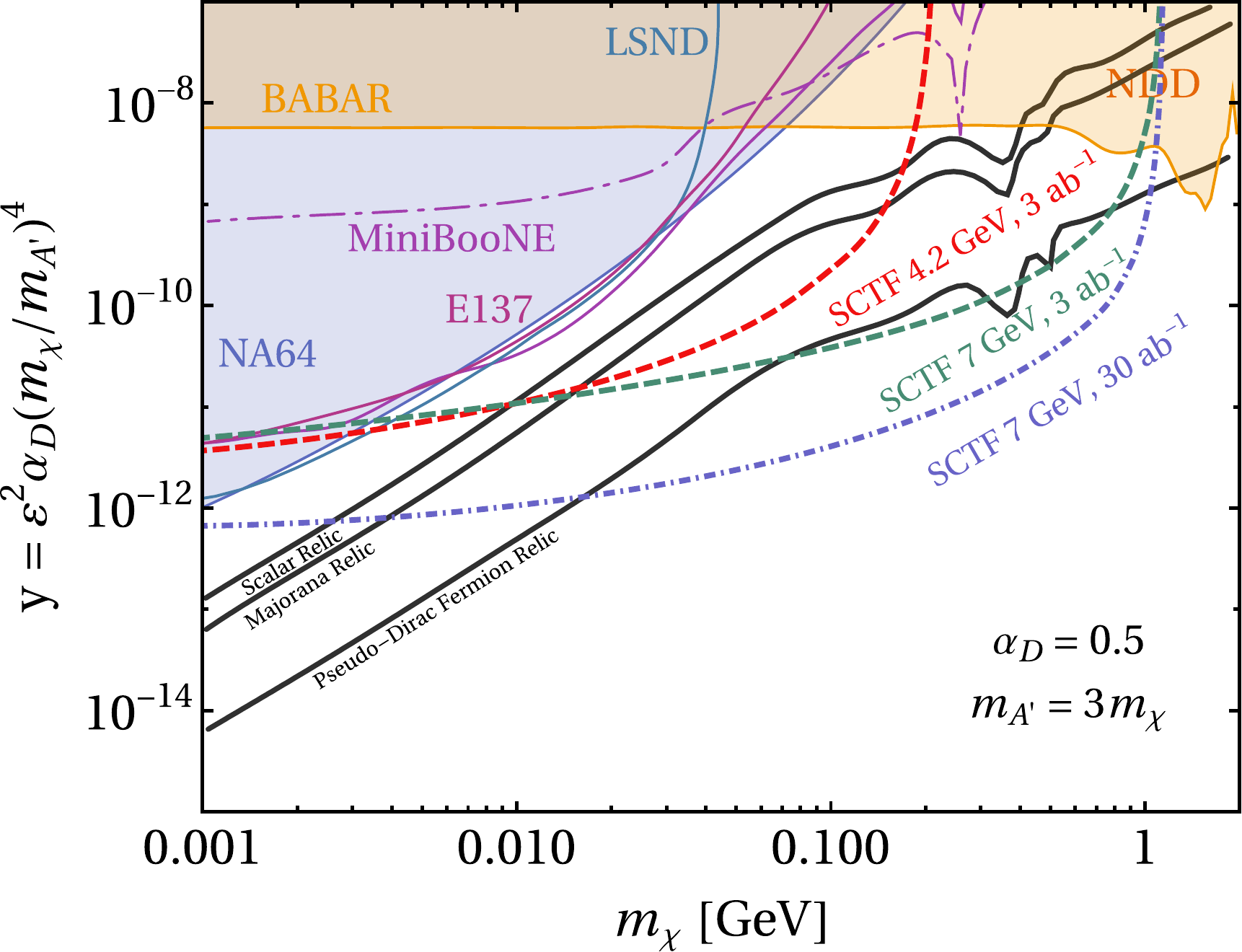}
  \hspace{0.3cm}
  \includegraphics[width=0.48\textwidth]{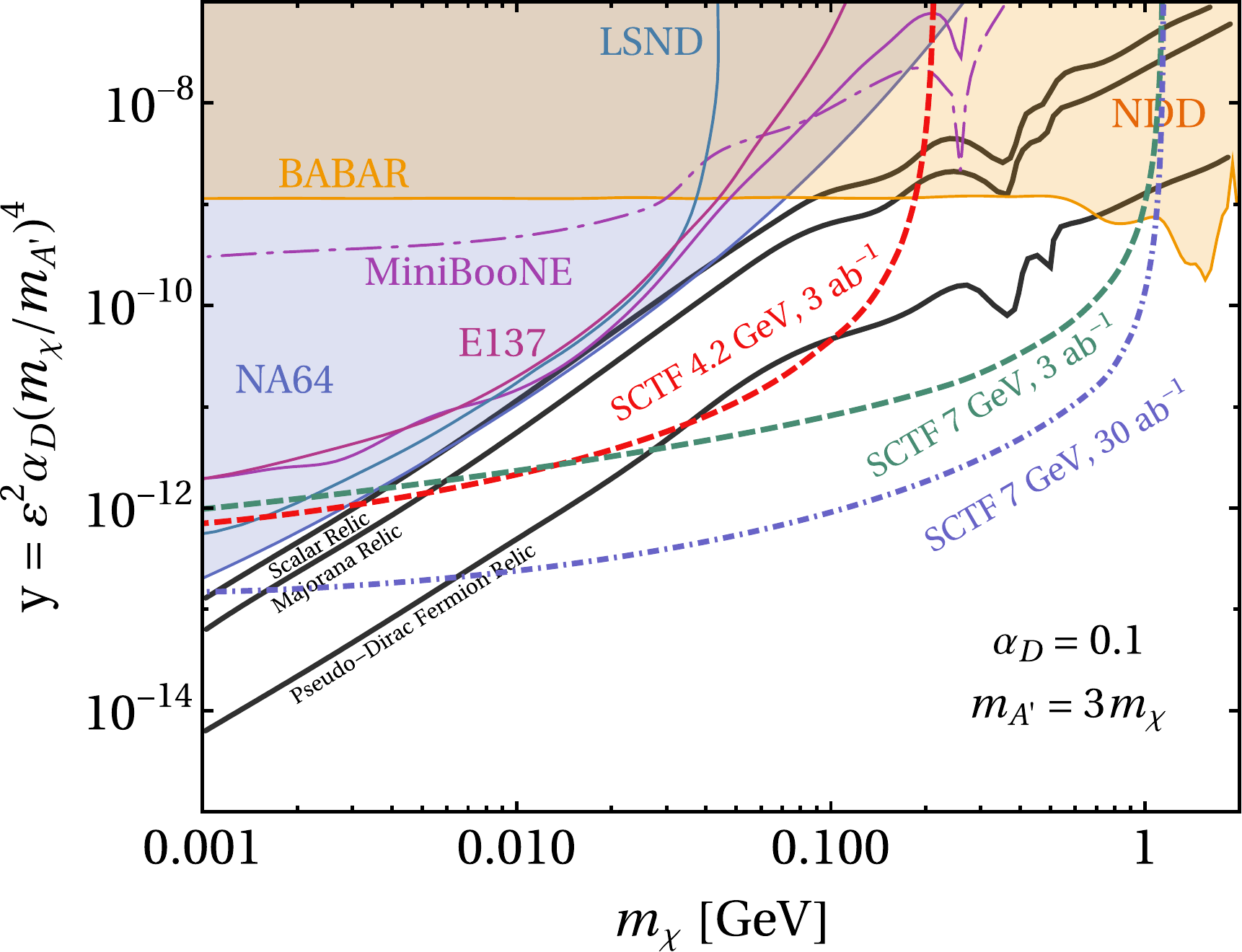}
\end{center}
\caption{The Sensitivity curves at the $90\%$ C.L. in the~$[y,m_{\chi}]$ plane are obtained for $\alpha_{\rm D}=0.5$ (left pannel) and 
$\alpha_{\rm D}=0.1$ (right panel) assuming integrated luminosity of $3\,{\rm ab}^{-1}$ 
at the collider energies $\sqrt{s}=4.2\,\GeV$ (dashed red) and $\sqrt{s}=7\,\GeV$ (dashed green), and assuming integrated luminosity of $30\,{\rm ab}^{-1}$ 
at $\sqrt{s}=7\,\GeV$ (dashed blue). 
The favored parameters values to account for the observed relic DM density for 
the scalar, pseudo-Dirac and Majorana type of light DM~\cite{Berlin:2018bsc} are shown as the solid lines.
The existing limits are shown in comparison with bounds obtained in Refs.~\cite{Banerjee:2019pds,Beacham:2019nyx,MiniBooNEDM:2018cxm,Battaglieri:2017aum,BaBar:2017tiz,Izaguirre:2017bqb,Alexander:2016aln,Izaguirre:2015yja,Izaguirre:2014bca,Batell:2014mga,Essig:2012yx,deNiverville:2011it,Batell:2009di}.}
\label{yy}
\end{figure*}

Using constraints on the cross section of the DM annihilation freeze-out and the obtained SCTF sensitivities on the mixing strength $\varepsilon$, 
it can be possible to derive constraints on the light thermal DM models.
In Fig.~\ref{yy} we plot the expected values at the $90\%$ C.L. for the dimensionless DM annihilation cross section parameter 
$y=\varepsilon^2\alpha_{\rm D}(m_{\chi}/\mAp)^4$ depending on the DM mass $m_{\x}$.
The limits on the variable $y$ are calculated under the convention $\alpha_{\rm D}=0.5$ and $\alpha_{\rm D}=0.1$,
and $\mAp=3 m_{\x}$~\cite{Beacham:2019nyx,Battaglieri:2017aum,Izaguirre:2015yja}, larger ratios $m_{\x}/\mAp$ qualitatively change the physics and larger $\alpha_{\rm D}$ can run towards the non-perturbative regime~\cite{Davoudiasl:2015hxa}.

The favored parameters for the scalar, pseudo-Dirac (with a small splitting) and Majorana light thermal DM scenarios, taking into account the observed DM relic density, shown as the solid black lines~\cite{Berlin:2018bsc}. Bounds from other experiments for comparison are also shown~\cite{Banerjee:2019pds,Beacham:2019nyx,MiniBooNEDM:2018cxm,Battaglieri:2017aum,BaBar:2017tiz,Izaguirre:2017bqb,Alexander:2016aln,Izaguirre:2015yja,Izaguirre:2014bca,Batell:2014mga,Essig:2012yx,deNiverville:2011it,Batell:2009di}.

We can see that the model for scalar DM can be excluded by the combined possible future data from the SCTF at $\sqrt{s}=7\,\GeV$, 
$3\,{\rm ab}^{-1}$ and $30\,{\rm ab}^{-1}$ and BABAR~\cite{BaBar:2017tiz}: for scalar DM in the mass region
$0.4\,\GeV\lesssim m_{\x}\lesssim 1\,\GeV$ and Majorana DM~--- 
$0.5\,\GeV\lesssim m_{\x}\lesssim 1\,\GeV$ by $\alpha_{\rm D}=0.5$, $0.1\,\GeV\lesssim m_{\x}\lesssim 1\,\GeV$ and
$0.2\,\GeV\lesssim m_{\x}\lesssim 1\,\GeV$ by $\alpha_{\rm D}=0.1$, respectively; 
for pseudo-Dirac DM in the mass region $0.7\,\GeV\lesssim m_{\x}\lesssim 1\,\GeV$ by $\alpha_{\rm D}=0.1$.

\begin{figure}[htb]
	\begin{center}
		\includegraphics[width=0.48\textwidth]{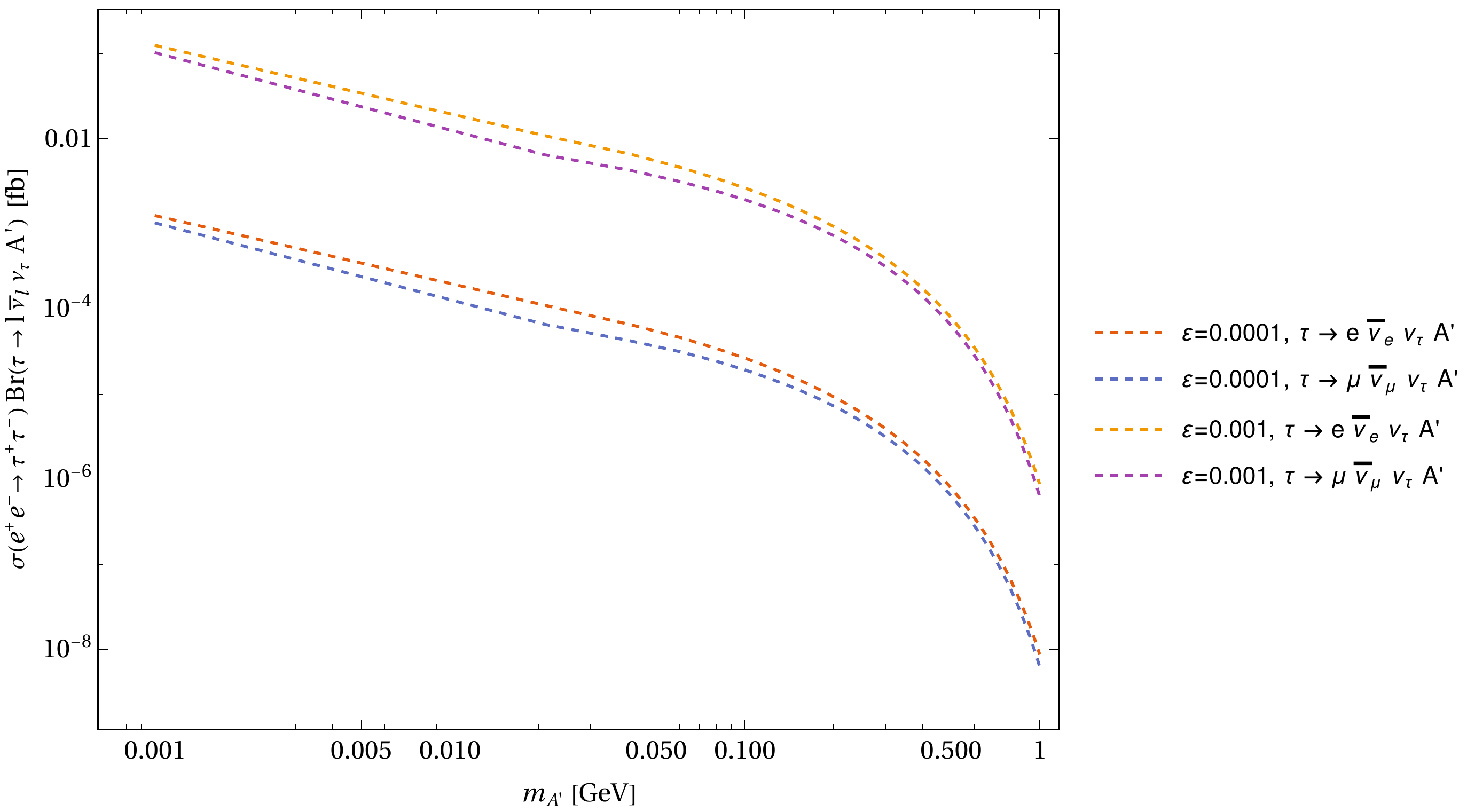}
	\end{center}
	\caption{The cross sections of $\Ap$ production in $\ee$ collisions at the energy of $4.2\,\GeV$ by tau decays in the dominant decay modes 
		$\tau\ra\ell\bar{\nu}_{\ell}\nu_{\tau}\Ap$, 
		$\ell=e, \mu$.}
	\label{tau-decay}
\end{figure}

In addition to the processes of associative production of a vector mediator with $\tau$ leptons, we evaluated the possibility of detecting mediators in $\tau$ decays $\tau\ra\ell\bar{\nu}_{\ell}\nu_{\tau}\Ap$, where
$\ell=e, \mu$.
To estimate, we used the SCTF running energy of $4.2\,\GeV$ with the highest value of the $\tau$ pair production cross section.
The Fig.~\ref{tau-decay} shows the dependence of these processes cross sections for $\varepsilon=0.0001$ and $\varepsilon=0.001$. It can be seen that taking into account the modes of the collider operation, one should not expect the appearance of such events for 
$\epsilon\lesssim 0.001$ in the $\MeV-\GeV$ mass of $\Ap$ range.

\section{Summary}

\begin{figure*}[htb]
	\begin{center}
		\includegraphics[width=0.48\textwidth]{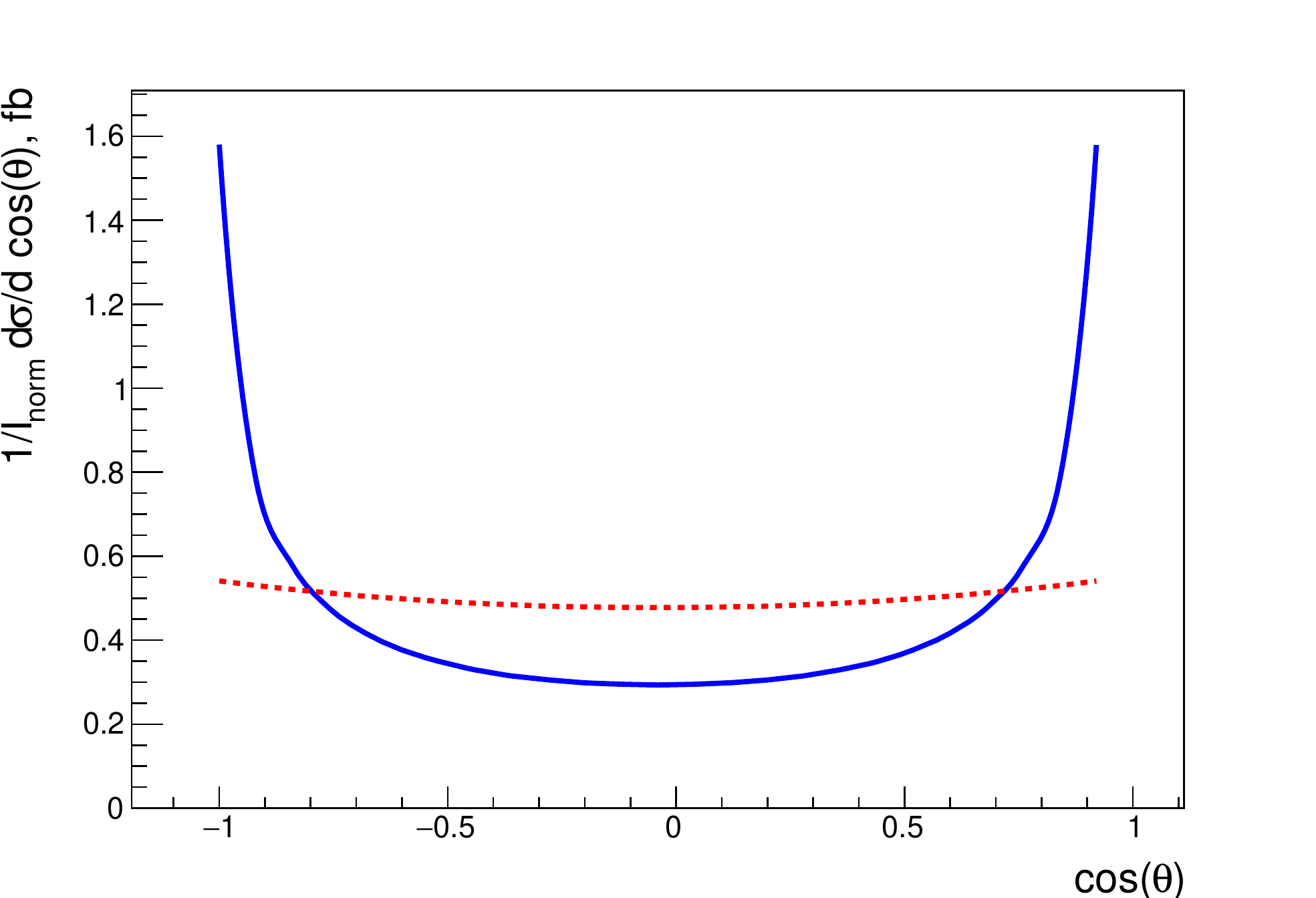}
		\hspace{0.3cm}
		\includegraphics[width=0.48\textwidth]{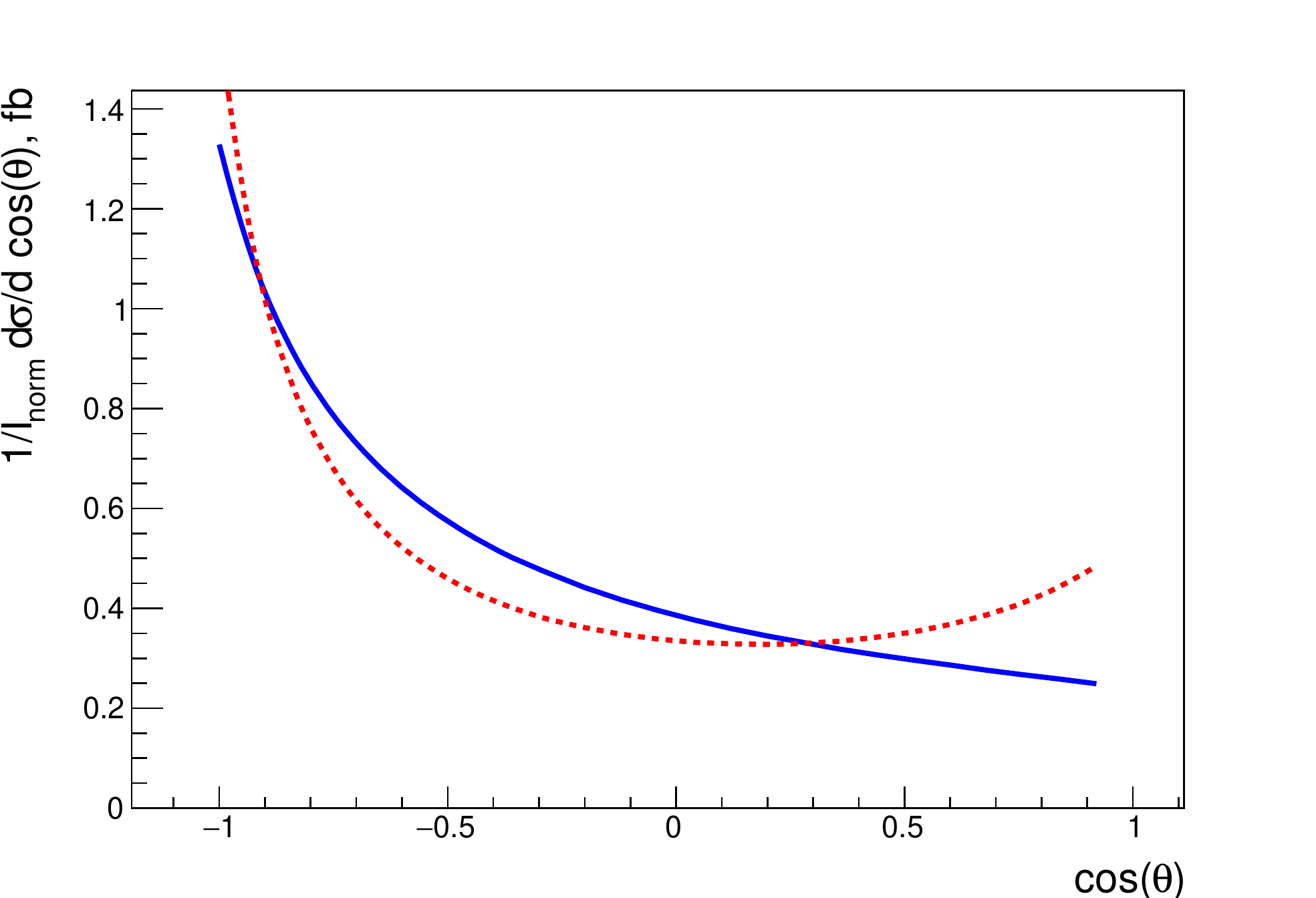}
	\end{center}
	\caption{Normalized $\cos\theta$ distribution. On the left plot: $\theta$ is an angle between momentums of initial electron and final mediator. On the right plot: $\theta$ is an angle between momentums of final tau and final mediator. 
		Red doted line correspond to dark scalar with mass = $0.5\,\GeV$ and $\xi=0.1$
	 and blue solid line correspond to dark photon with $\varepsilon=0.0003$. $\sqrt{s}=7\,GeV$.}
	\label{ad-15-7}
\end{figure*}

In this work we have proposed a search for invisible decays of dark leptophilic scalar and dark photon in the processes 
$\ee\ra\tautau\pL$ and $\ee\ra\tautau\Ap$, respectively, at the future SCTF. 
We present the promising sensitivity on the scalar coupling constant $\xi$ and the dark photon mixing strength $\varepsilon$ at $\sqrt{s}=4.2,\, 7\,\GeV$ assuming integrated luminosity of $3\,{\rm ab}^{-1}$, and $30\,{\rm ab}^{-1}$ for $\sqrt{s}=7\,\GeV$ in the non-excluded parameter spaces below $4\,\GeV$. 
In addition, we have discussed the constraints on light thermal DM. 
We provide the expected $90\%$~C.L. sensitivity on the dimensionless DM annihilation cross section parameter $y$. 
We find that the future SCTF data can expand the search light DM for the mass region $0.001\,\GeV\lesssim\x\lesssim 1\,\GeV$; for the scalar, pseudo-Dirac and Majorana type of light DM can be excluded by the combined data from the future STCF and BABAR for the mass region $0.1\,\GeV\lesssim\x\lesssim 1\,\GeV$.

It should be noted that the search for light dark matter in the processes of associated production with tau leptons is of particular interest, since it allows one to simultaneously search for scalar and vector mediators. As shown in Fig.~\ref{ad-15-7}, the angular distributions differ significantly for mediators with different spins. By studying the angular correlations in such processes, one can determine the spin nature of mediator particles.

\section*{Acknowledgments}
The study was carried out within the framework of the scientific program of the National Center for Physics and Mathematics, the project "Particle Physics and Cosmology" and was supported by the Foundation for the Advancement of Theoretical Physics and Mathematics ``BASIS''.

\end{document}